\documentclass[pra,twocolumn,showpacs,amssymb,aps]{revtex4}
\usepackage{dcolumn}
\usepackage{bm}
\usepackage{graphicx}
\usepackage{epstopdf}
\usepackage{subfig}
\usepackage{color}
\usepackage{amsmath}
\usepackage{tikz-cd}
\usepackage[colorlinks=true]{hyperref}
\newcommand*\sech{\mathop{}\!\operatorname{sech}}
\newcommand*\cn{\mathop{}\!\operatorname{cn}}
\newcommand*\sn{\mathop{}\!\operatorname{sn}}
\newcommand*\dn{\mathop{}\!\operatorname{dn}}
\begin{document}
	\title{Doubly-Periodic Solutions of the Class I Infinitely Extended Nonlinear Schr\"odinger Equation}
	\author{M. Crabb and N. Akhmediev}
	\affiliation{Optical Sciences Group, Research School of Physics and Engineering, The Australian National University, Canberra, ACT, 2600, Australia}
	\begin{abstract} 
	We present doubly-periodic solutions of the infinitely extended nonlinear Schr\"odinger equation with an arbitrary number of higher-order terms
	and corresponding free real parameters. Solutions have one additional free variable parameter that allows to vary periods along the two axes. The presence of infinitely many free parameters provides many possibilities in applying the solutions to nonlinear wave evolution.
		Being general, this solution admits several particular cases which are also given in this work.
	\end{abstract}
	\maketitle
	\date{today}
\section{Introduction}
Evolution equations are a powerful tool for describing a great variety of physical effects. These include pulse propagation in optical fibers \cite{Agrawal,Dudley1}, nonlinear ocean wave phenomena \cite{Osborne,Grimshaw}, plasma \cite{Swanson,Kolomeisky} and atmospheric \cite{Gossard,Stenflo} waves, and the dynamics of Bose-Einstein condensates \cite{Pitaevskii,Andreev,Bobrov} to mention only a few. Using evolution equations, one can explain phenomena that would otherwise be difficult to interpret. Examples of such phenomena include solitons \cite{Zabusky,Gerdjikov}, modulation instability \cite{BF,Trillo,Mussot12}, supercontinuum generation \cite{Dudley}, Fermi-Pasta-Ulam recurrence \cite{Mussot}, rogue waves \cite{Baronio,Onorato,Baronio3,KPS,ds-oyI}, etc. It is especially helpful when the evolution equations under study are integrable \cite{Ablowitz}.
Unfortunately, this is not always the case. Not all evolution equations are integrable \cite{Sakovich,Slunyaev,Sedlet}. Finding new integrable equations \cite{Porsezian97,Marchant,Baronio2013}, and extending the existing ones to allow for incorporating new, physically relevant terms \cite{porsezian1992,Mihalache2,Mihalache3,Trippenbach,Matsuno80}, is therefore an important direction of research in nonlinear dynamics.\\\indent	
The  nonlinear Schr\"{o}dinger  equation (NLSE) is one of the fundamental examples of a completely integrable equation \cite{Zakharov,bk} which finds application in the description of water waves \cite{Zakharov2}, pulses in optical fibres \cite{Hasegawa}, amongst other areas of physics. In neither of these fields is the NLSE absolutely accurate. Extensions of the NLSE that have physical relevance are therefore essential, and these have been considered in a number of works that include both optical applications and water waves \cite{Slunyaev,Sedlet}.\\\indent
In general, these extensions lift the integrability for most particular physical problems. However, in special cases, we can obtain extensions which remain integrable, and, in addition, we can add infinitely many higher-order terms with variable coefficients representing the strength of these effects, adding substantial flexibility to the evolution equation.\\\indent
There are two types of extensions of the NLSE \cite{Chaos2015,PRE16,PRE17,Naturforsch2018,CHAOS2018}. For clarity, we call them here the class I and class II extensions. The next higher-order terms in the class I extension correspond to the Hirota equation  \cite{Chaos2015,PRE16,PRE17}, while the next higher-order terms in the class II extension correspond to the Sasa-Satsuma equation \cite{Naturforsch2018,CHAOS2018}.\\\indent
Both of these extensions take into account higher order dispersive effects, without any restriction on the magnitude of these effects, i.e. without requiring them to be small perturbations. In practice, waves are affected by more than just second-order dispersion, so solutions to the infinite equations are an important development in that they allow a generalisation of the fundamental structures which appear in the `basic' nonlinear Schr\"odinger equation to account for these effects. When the number of higher-order terms is limited to the third order, integrability can be achieved with variation of two free parameters \cite{Mihalache2019}. For infinitely extended equations, the number of free parameters is also infinite.\\\indent
The presence of two classes of integrable extensions thus widens the range of problems that can be solved analytically. Remarkably, solutions to both classes can be found in general form, even for the case of an infinite number of terms, and an infinite number of corresponding parameters. In order to find these solutions, we can start with the known solutions of the NLSE and extend them, recalculating the parameters of the solution. This can be done for soliton, breather and rogue wave solutions \cite{PRE16,PRE17}. In the present work, we further expand this approach to doubly-periodic solutions. They include as particular cases solitons and breathers.\\\indent
To be specific, we start with the standard focusing NLSE:
\begin{equation}\label{basic}
i\frac{\partial\psi}{\partial x}+\alpha_2\left(\frac{\partial^2\psi}{\partial t^2}+2|\psi|^2\psi\right)=0
\end{equation} 
where $\psi=\psi(x,t)$ is the wave envelope, $x$ is the distance along the fibre or along the water surface, while $t$ is the retarded time in the frame moving with the group velocity of wave packets. The coefficient $\alpha_2$ scales the dispersion and nonlinear terms in a way convenient for the extensions. 
The infinite extension of Eq. (\ref{basic}) can be written in the form
\begin{equation}\label{hierarchy}
i\frac{\partial\psi}{\partial x}+\sum_{n=1}^{\infty}(\alpha_{2n}K_{2n}[\psi]-i\alpha_{2n+1}K_{2n+1}[\psi])=0,
\end{equation}
where the $K_n[\psi]$ are $n$th order differential operators, and the coefficients $\alpha_n$ are arbitrary real numbers.\\\indent
Here we deal with the class I extension, and exact forms for the class I form of the operators $K_n[\psi]$ are given in \cite{PRE16}. The four lowest order operators $K_n$ are presented below.
\begin{eqnarray}\label{mkdv}
K_2[\psi] &=& \psi_{tt}+ 2  |\psi|^2   \psi   ,
\\  \nonumber
K_3[\psi] &=& \psi_{ttt}+6  |\psi|^2   \psi_t   ,
\\  \nonumber
K_4[\psi]  &=&  \psi _{tttt} + 8 |\psi|^2 \psi_{tt}   +6 \psi   |\psi|^4+
\\  \nonumber
&& + 4 \psi  | \psi_t|^2 + 6 \psi_t ^2 \psi^*  + 2 \psi ^2
   \psi^*_{tt}. 
\\   \nonumber
K_5[\psi]  &=&   \psi _{ttttt} +10|\psi|^2  \psi_{ttt}+10(\psi\,|\psi_{t}|^2)_t+
\\   \nonumber
&& +   20 \psi^*  \psi_t \psi _{tt}+30|\psi|^4 \psi_{t}.
\end{eqnarray}  
The coefficients $\alpha_n$ determine the strength of the dispersive effects of order $n$, as well as higher-order nonlinear effects. The whole infinite equation (\ref{hierarchy}) is integrable for arbitrary values of $\alpha_n$.  
For example, the equation with the terms up to the third order is the Hirota equation
\begin{equation}\label{hirota}
i\frac{\partial\psi}{\partial x} + \alpha_2 \left( \frac{\partial^2\psi}{\partial t^2} + |\psi|^2\psi \right)-i\alpha_3 \left( \frac{\partial^3 \psi}{\partial t^3} + 6|\psi|^2\frac{\partial\psi}{\partial t} \right)=0
\end{equation}
while including up to fourth order terms gives the Lakshmanan-Porsezian-Daniel (LPD) equation  \cite{porsezian1992}, and so on. Particular solutions of the first-order to the equation (\ref{hierarchy}) have been given in \cite{PRE16,PRE17}.  Solutions of  the mKdV equation, which is a particular case of (\ref{hierarchy}), are provided in \cite{mkdv}. Thus, any of the extensions of (\ref{hierarchy}) with only a few nonzero terms can be considered individually.\\\indent
Among the more general families of solutions to the nonlinear Schr\"odinger equation (\ref{basic}) are the doubly-periodic solutions \cite{AEK}. The two periods of this family can be varied, thus providing particular cases in the form of solitons, breathers, cnoidal, dnoidal and Peregrine waves when one or two of these periods tend to infinity or zero \cite{AEK}.
 Unlike breather solutions, however, these doubly-periodic solutions do not decay in either space or time, and instead have the special property of being periodic in both the $x$ and $t$ variables.\\\indent 
In this work, we show that doubly periodic solutions can be found for the class I equation. This solution includes infinitely many parameters $\alpha_n$ in full generality. We  also show that particular limiting cases of this family include the Akhmediev breather and soliton solutions.
% Section 2
 \section{Doubly Periodic solutions}
There are two types of doubly periodic solutions to the nonlinear Schr\"odinger equation, which can be classified as type-A and type-B \cite{adiabatic}. Each of them is expressed in terms of Jacobi elliptic functions, with the modulus $k$ as the free parameter of the family. First, we consider the type-A solutions.
% subsection
\subsection{Type-A Solutions}
\begin{figure}[ht]
	\includegraphics[scale=0.28]{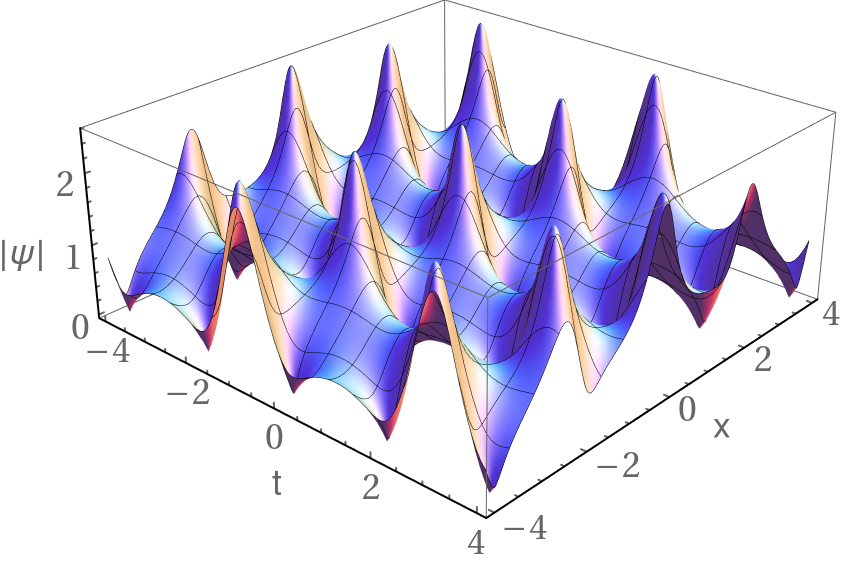}
	\caption{\it Type-A solution for Eq. (\ref{hierarchy}), with $k=0.7,$ $\alpha_n=1/n!$ up to $n=10$ and all other $\alpha_n=0.$ Notice that $v$ is nonzero.}
	\label{DoubleTypeA}
\end{figure}

Type-A solutions of Eq. (\ref{hierarchy}) are of the form
\begin{equation}\label{typeA}
\psi(x,t)=\frac{k\operatorname{sn}\left(Bx/k,k\right)-iC(t+vx)\operatorname{dn}\left(Bx/k,k\right)}{k-kC(t+vx)\operatorname{cn}\left(Bx/k,k\right)}e^{i\phi x}
\end{equation}
where
\[C(t)=\sqrt{\frac{k}{1+k}}\operatorname{cn}\left(\sqrt{\frac{2}{k}}t,\sqrt{\frac{1-k}{2}}\right).\]
The constants $B$, $v$ and $\phi$ in the solution (\ref{typeA}) are given in terms of the coefficients $\alpha_n$ of the equation (\ref{hierarchy}). Taking into account the lowest order terms, step by step, we find:
\begin{eqnarray}
B &=& 2\alpha_2+8\alpha_4+\left(32-\frac{4}{k^2}\right)\alpha_6   \nonumber\\
& + & \left(128-\frac{32}{k^2}\right)\alpha_8
+\left(512-\frac{192}{k^2}+\frac{12}{k^4}\right)\alpha_{10}   \nonumber \\ 
&+& \left(2048-\frac{1024}{k^2}+\frac{144}{k^4}\right)\alpha_{12} +\cdots \\
\phi &=& 2\alpha_2+\left(8-\frac{2}{k^2}\right)\alpha_4+\left(32-\frac{12}{k^2}\right)\alpha_6  \nonumber\\
& +& \left(128-\frac{64}{k^2}+\frac{6}{k^4}\right)\alpha_8+\left(512-\frac{320}{k^2}+\frac{60}{k^4}\right)\alpha_{10}\nonumber\\
& +& \left(2048-\frac{1536}{k^2}+\frac{432}{k^4}-\frac{20}{k^6}\right)\alpha_{12}+\cdots\\
v&=&4\alpha_3+\left(16-\frac{2}{k^2}\right)\alpha_5+\left(64-\frac{16}{k^2}\right)\alpha_7 \nonumber\\
&+&\left(256-\frac{96}{k^2}+\frac{6}{k^4}\right)\alpha_9+\left(1024-\frac{512}{k^2}+\frac{72}{k^4}\right)\alpha_{11} \nonumber\\
&+&\left(4096-\frac{2560}{k^2}+\frac{576}{k^4}-\frac{20}{k^6}\right)\alpha_{13}+\cdots
\end{eqnarray}\indent
An important observation here is that the expression for $v$ which is responsible for the `tilt' in $(x,t)$-plane discussed below includes only odd-order coefficients $\alpha_n$. If these are zero, $v$ is also zero.\\\indent
In order to determine the general forms, with all $\alpha_n$ included, we note that only one of these sets of polynomial coefficients is algebraically independent. 
The coefficient of $\alpha_{2n-1}$ in $v$ is half the coefficient of $\alpha_{2n}$ in $B,$ and is also the sum of the coefficients of $\alpha_{2n}$ in $B$ and $\phi.$ It is therefore sufficient to determine the coefficients of $B,$ since, if we let
	\[B=\sum_{n=1}^{\infty}B_n\alpha_{2n},\]
	then
	\[v=\sum_{n=1}^{\infty}\tfrac12B_{n+1}\alpha_{2n+1},~\phi=\sum_{n=1}^{\infty}(\tfrac12B_{n+1}-B_n)\alpha_{2n}.\]
Calculating further the other terms of $B_n$, we find that they are the polynomials
	%\[2^{2n-1}\left\{1-\frac{n-2}{2}\frac{1}{4k^2}+3\frac{(n-3)(n-4)}{2^2}\frac{1}{16k^4}-\cdots\right\},\]
	\[B_n=2^{2n-1}\sum_{r=0}^{\lfloor\tfrac12n\rfloor}\frac{(-1)^r(2r)!(n-r-1)!}{2^{4r}(r!)^3(n-2r-1)!}\frac{1}{k^{2r}},\]
where the summation ends at $\lfloor\tfrac12n\rfloor$ terms, $\lfloor\cdot\rfloor$ being the floor function, i.e. $\lfloor m \rfloor$ is the largest integer which is not greater than $m$. Now the full expression for $B$ is given explicitly by the series formula
	\begin{equation}B=\sum_{n=1}^{\infty}2^{2n-1}\sum_{r=0}^{\lfloor\tfrac12n\rfloor}\frac{(-1)^r}{2^{4r}}\binom{2r}{r}\binom{n-r-1}{r}\frac{1}{k^{2r}}\alpha_{2n}.\end{equation}
	It also follows that
	\begin{equation}v=\sum_{n=1}^{\infty}2^{2n}\sum_{r=0}^{\lfloor\tfrac12(n+1)\rfloor}\frac{(-1)^r}{2^{4r}}\binom{2r}{r}\binom{n-r}{r}\frac{1}{k^{2r}}\alpha_{2n+1},\end{equation}
and the phase factor $\phi$ is
	\begin{align}\phi&=\sum_{n=1}^{\infty}2^{2n-1}\bigg\{2\sum_{r=0}^{\lfloor\tfrac12(n+1)\rfloor}\frac{(-1)^r}{2^{4r}}\binom{2r}{r}\binom{n-r}{r}\frac{1}{k^{2r}}-\nonumber\\
	&~~~-\sum_{r=0}^{\lfloor\tfrac12n\rfloor}\frac{(-1)^r}{2^{4r}}\binom{2r}{r}\binom{n-r-1}{r}\frac{1}{k^{2r}}\bigg\}\alpha_{2n}\end{align}\\\indent
We plot an example of the type-A solution for Eq. (\ref{hierarchy}) in Fig. \ref{DoubleTypeA}. For this example, we take the modulus $k=0.7$, and the coefficients $\alpha_n=1/n!$ up to $n=10$, restricting ourselves with the case when all terms higher than $n=10$ are zero.
% Figure 1
We can see from Fig. \ref{DoubleTypeA} that $v$ introduces a tilt to the solutions and appears to operate similarly to a velocity parameter in a boost transformation. However, note that $v$ cannot be interpreted exactly as a velocity, as there is no function $f$ such that we could write
$\psi(x,t)=f(t+vx)$
as we could do with a travelling wave, except in the case that $\phi=B=0.$\\\indent
From equation (\ref{typeA}), we can see that the parameter $B/k$ can be associated with a frequency of the modulation along the $x$ axis when $v=0$.
On the other hand, the real quarter-period along the $t$ axis is: 
\[\sqrt{\frac{k}{2}}K\left(\sqrt{\frac{1-k}{2}}\right),\]
where $K(k)$ denotes the complete elliptic integral of the first kind with modulus $k$. However, just as $v$ cannot be precisely interpreted as a velocity, neither can $B/k$ be thought of as a modulation frequency exactly, except when $v=0$ and the solution is periodic along the $x$-axis.

\subsection{The Akhmediev Breather Limit}
\begin{figure}[ht]
	\includegraphics[scale=0.33]{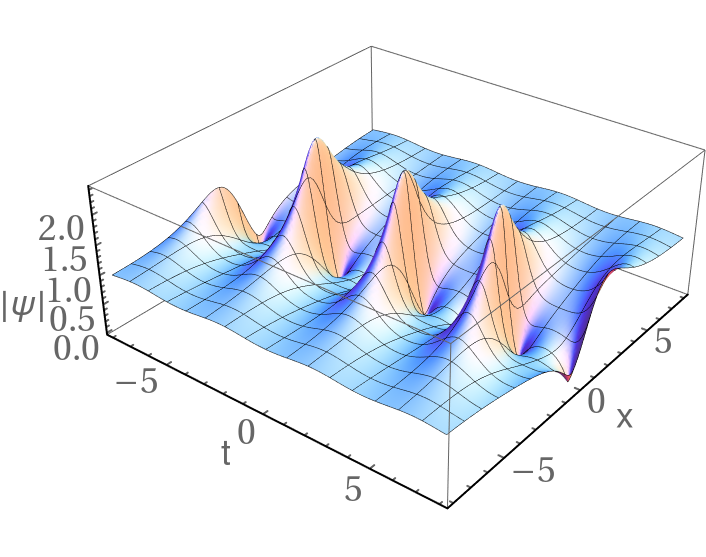}
	\caption{\it  {The limiting case $k\to1$ of the type-A solutions, with $\alpha_n=(n!)^2/(2n)!$ up to $n=12,$ all other $\alpha_n=0$. This is the Akhmediev breather solution of Eq. (\ref{hierarchy}), periodic in $t$ and with the growth-decay cycle in $x$.}}
	\label{AB}
\end{figure}

In the limit as modulus $k\to1,$ we have
\begin{align}
\label{Ba1}
	\lim\limits_{k\to1}B&=\sum_{n=1}^{\infty}\binom{2n}{n}nF(1-n,1;\tfrac32;\tfrac12)\alpha_{2n},\\
\label{fa1}
	\lim\limits_{k\to1}\phi&=\sum_{n=1}^{\infty}\binom{2n}{n}\alpha_{2n}\\
\label{va1}
	\lim\limits_{k\to1}v&=\sum_{n=1}^{\infty}\binom{2n}{n}(2n+1)F(-n,1;\tfrac32;\tfrac12)\alpha_{2n+1}
\end{align}
where $F(a,b;c;z)$ is Gauss' hypergeometric function. The type-A solution then reduces to the Akhmediev breather with modulation parameter $\sqrt2$ \cite{PRE16}; i.e. the solution becomes
\begin{equation}
\label{ab}
\lim\limits_{k\to1}\psi(x,t)=\frac{\sqrt2\sinh Bx-i\cos\sqrt2(t+vx)}{\sqrt2\cosh Bx-\cos\sqrt2(t+vx)}e^{i\phi x}\end{equation}
with $B,$ $\phi$, and $v$ given by (\ref{Ba1}), (\ref{fa1}), and (\ref{va1}), respectively. An example of this limiting case is plotted in Fig. \ref{AB}.
% Figure 2

% subsection
\subsection{Type-B Solutions}
\begin{figure}[h!]
	\includegraphics[scale=0.28]{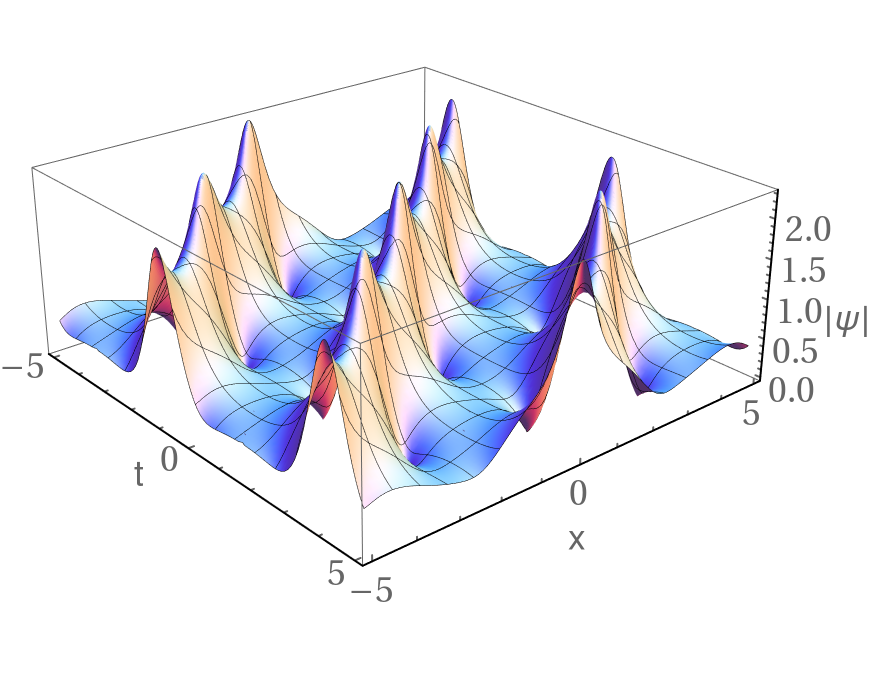}
	\caption{\it The type-B solution (\ref{typeb}), where $\kappa=0.7,$ $\alpha_n=1/n!$ up to $n=8,$ with all other $\alpha_n=0$. The peaks of this solution are aligned along lines of constant $\tau=t+vx.$}
	\label{TypeB}
\end{figure}

Type-B solutions can be considered as the analytic continuation of the type-A solutions for values of the modulus $k>1$. Using the corresponding transformations of the elliptic functions \cite{Gradshtein} with modulus $\kappa=1/k$, these solutions take the form
\begin{equation}
\label{typeb}
	\psi(x,t)=\kappa e^{i\phi x}\frac{\sqrt{1+\kappa}\sn(Bx,\kappa)-iA(t+vx)\cn(Bx,\kappa)}{\sqrt{1+\kappa}-A(t+vx)\dn(Bx,\kappa)}
\end{equation}
where the function $A(t)$ is given by
\[A(t)=\operatorname{cd}\left(\sqrt{1+\kappa}t,\sqrt{\frac{1-\kappa}{1+\kappa}}\right),\]
and $\kappa$ is in the range $0<\kappa<1.$ 
In this case, we find:
\begin{align}
B&=\sum_{n=1}^{\infty}2^{2n-1}\sum_{r=0}^{\lfloor\tfrac12n\rfloor}\frac{(-1)^r}{2^{4r}}\binom{2r}{r}\binom{n-r-1}{r}\kappa^{2r}\alpha_{2n},\\
\phi&=\sum_{n=1}^{\infty}2^{2n-1}\bigg\{2\sum_{r=0}^{\lfloor\tfrac12(n+1)\rfloor}\frac{(-1)^r}{2^{4r}}\binom{2r}{r}\binom{n-r}{r}\kappa^{2r}-\nonumber\\
&~~~-\sum_{r=0}^{\lfloor\tfrac12n\rfloor}\frac{(-1)^r}{2^{4r}}\binom{2r}{r}\binom{n-r-1}{r}\kappa^{2r}\bigg\}\alpha_{2n}\\
v&=\sum_{n=1}^{\infty}2^{2n}\sum_{r=0}^{\lfloor\tfrac12(n+1)\rfloor}\frac{(-1)^r}{2^{4r}}\binom{2r}{r}\binom{n-r}{r}\kappa^{2r}\alpha_{2n+1}.
\end{align}
These are just the same polynomials as previously given for the type-A solutions, but with reciprocal argument $\kappa=1/k$.\\\indent
We plot an example of type-B solutions in Fig. \ref{TypeB}. The solution is qualitatively different from the type-A solution as the location of maxima are now different.
\\\indent
The limit as $\kappa\to1$ is identical to the limit as $k\to1$ for the reasons just stated previously, and we again recover the separatrix  breather solution (\ref{ab}). However, by changing the parameter $\kappa$, we can vary the periods of the type-B solutions while always keeping the functions analytic, so that in the limit $\kappa\to0$ we recover the soliton solution
\begin{equation}
	\lim\limits_{\kappa\to0}\psi(x,t)=2e^{i\phi x}\sech(2t+vx)
\end{equation}
with
\[v=\sum_{n=1}^{\infty}2^{2n-1}\alpha_{2n+1}\]
and
\[\phi=\sum_{n=1}^{\infty}2^{2n}\alpha_{2n},\]
which is the general soliton solution for the equation (\ref{hierarchy}), up to scaling \cite{PRE16}.

\section{Phase portrait of solutions}

\begin{figure}[ht]
\includegraphics[scale=0.15]{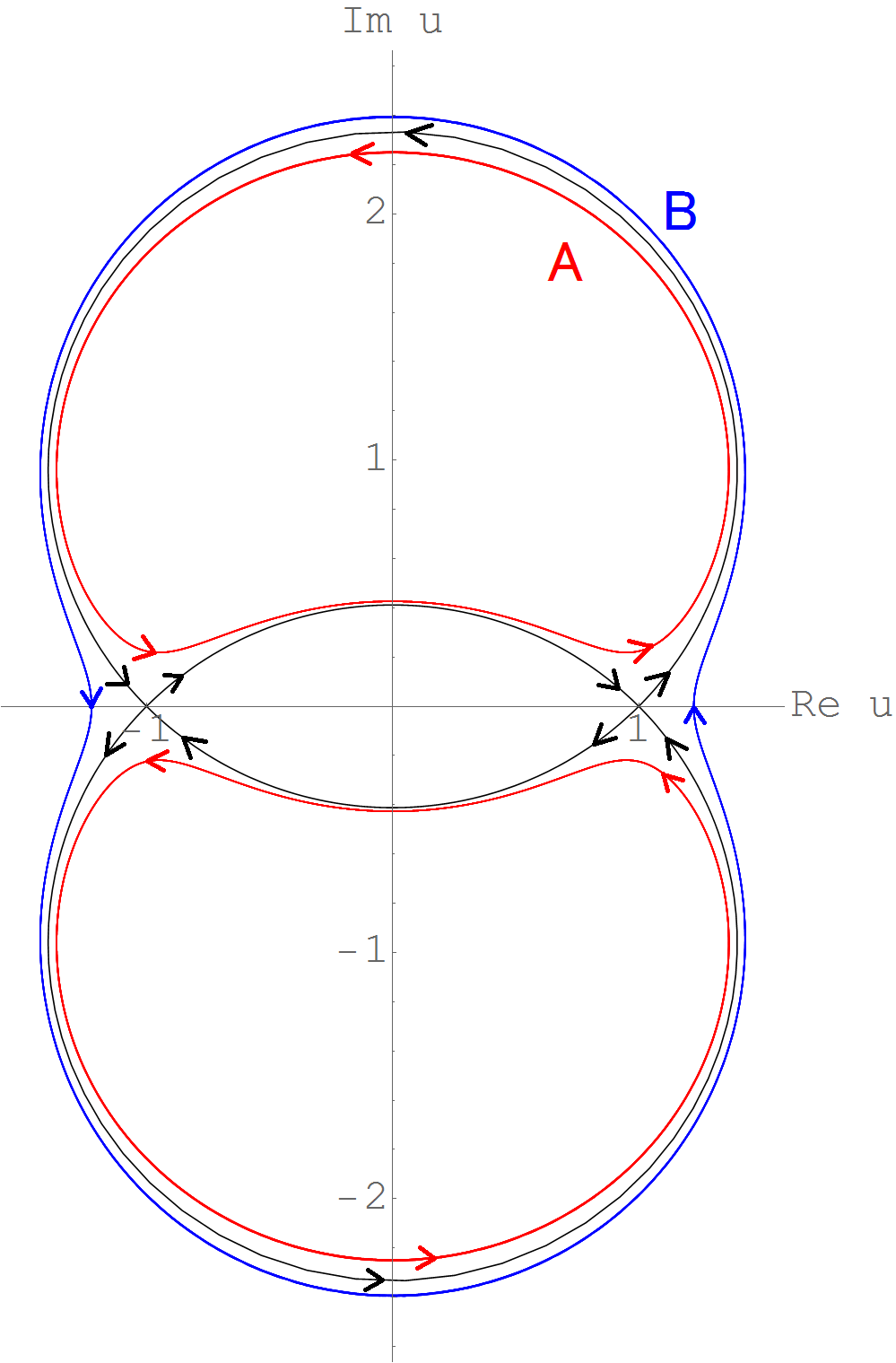}
\caption{\it  The phase portrait of the periodic dynamics around the Akhmediev breather shown by the black curves. The type-A solution is shown by the red curve while the type-B solution is shown by the blue curve. Here $u$ is defined by (\ref{u}), and the two saddle points are at $u=\pm1$. The trajectories are drawn along the lines $\xi=Bx$ and $\tau=t+vx$. Thus, evolution is in $\xi$ along lines of constant $\tau.$}
	\label{pp}
\end{figure}

The transformation of the two periodic solutions into the Akhmediev breather (AB) when $k\to1$ and $\kappa\to1$ can be illustrated by the phase portrait of these solutions which is shown in Fig. \ref{pp}.
Although the dynamical system that we are dealing with is infinite-dimensional, the dynamics of the solutions still can be presented on a two-dimensional plane which can be considered as a projection of the infinite-dimensional phase space onto a plane. The Akhmediev breather solution (\ref{ab}) on this plane is represented by the heteroclinic orbits connecting two saddle-points. It is a separatrix between the type-A and type-B solutions, represented respectively by the periodic orbits A and B.

% Figure 3

The projection of infinite-dimensional phase space onto a plane requires certain tricks, as the solutions involve drift. In order to avoid the corresponding shifts, we make the change of variable $\xi=Bx,$ $\tau=t+vx,$ and define the new function
\begin{equation}
\label{u}
u(\xi,\tau)=\psi(x,t)e^{-i\phi x}.
\end{equation}
The counterbalancing exponential factor allows us to stop the rotation of $\psi$ around the origin in the complex plane. Then it is easy to see that the trajectory corresponding to the AB solution satisfies the equation
\begin{equation}\label{circ}
	\{\operatorname{Re}u(\xi,T)\}^2+\{\operatorname{Im}u(\xi,T)-1\}^2=2,~T=\frac{n\pi}{\sqrt2},
\end{equation}
where $\operatorname{Re}u$ and $\operatorname{Im}u$ are the real and imaginary parts of $u,$ respectively, and $n$ is any integer. 
The trajectories defined by Eq. (\ref{circ}) are circular so long as we trace the evolution in $\xi$ along these lines of constant $\tau$. They are shown as black curves in Fig. \ref{pp}.
Similar precautions should be taken for the doubly-periodic orbits.\\\indent
The difference between the type-A and type-B solutions can be seen clearly from Fig. \ref{pp}.
Trajectories for the type-A solutions never cross the real axis, whereas trajectories for the type-B solutions do.
Therefore, each time they complete one full path, the phase change is either zero or $2\pi.$ The periodicity of solutions along the $x$-axis depends on the strength of the dispersion, or the values of the coefficients $\alpha_n$, through $\phi,$ $B$, and $v$. 
\section{The Case With Zero Even Order Terms}
\begin{figure}[ht]
	\includegraphics[scale=0.3]{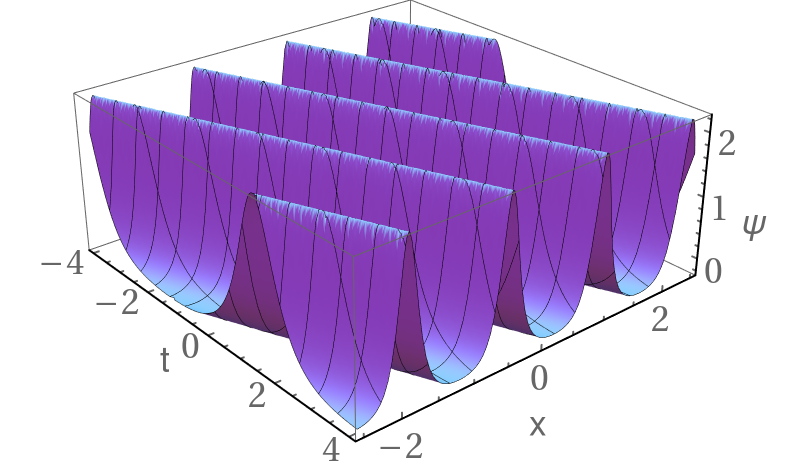}
	\caption{\it{
			Plot of real-valued mKdV equation solution $\psi$ given by
			(\ref{u1}), where $k=\tfrac12$. Here the type-B solution (\ref{typeb}) reduces to a periodic solution propagating with speed $v=4.$}}
	\label{periodicmkdv}
\end{figure}

In the absence of any even order terms in Eq. (\ref{hierarchy}), it becomes real:
\begin{equation}\label{real}
\frac{\partial\psi}{\partial x}-\sum_{n=1}^{\infty}\alpha_{2n+1}K_{2n+1}[\psi]=0,
\end{equation}
Then we have $\phi=0$ and $B=0$, and the type-B solution takes the real-valued form 
\begin{equation}
	\psi(x,t)=u(\tau)=\frac{\kappa A(\tau)}{\sqrt{1+\kappa}-A(\tau)},
\end{equation}
where $\tau=t+vx.$ Note that here $v$ can be interpreted as a velocity since $u$ has the form of a travelling wave.\\\indent
The real quarter-period in $\tau$ is equal to the real quarter-period in $t$ of the usual type-B solutions, which is
\[\frac{1}{\sqrt{1+\kappa}}K\left(\sqrt{\frac{1-\kappa}{1+\kappa}}\right).\]The real quarter-period in $x$, for fixed $t,$ is $\sqrt{\kappa}K(\kappa)/v$ since $\phi=B=0.$\\\indent
In particular, with the normalisation $\alpha_3=-1$ and all other $\alpha_n=0,$ this becomes the periodic solution to the mKdV equation 
$$\psi_x+\psi_{ttt}+6\psi^2\psi_t=0,$$
given by
\begin{equation}
\label{u1}
\psi(x,t)=\frac{\kappa\operatorname{cd}\left(\sqrt{1+\kappa}(t-4x),\sqrt{\frac{1-\kappa}{1+\kappa}}\right)}{\sqrt{1+\kappa}-\operatorname{cd}\left(\sqrt{1+\kappa}(t-4x),\sqrt{\frac{1-\kappa}{1+\kappa}}\right)}.
\end{equation}
We plot an example of this solution in Fig. \ref{periodicmkdv}.

% Figure 03
The degree of generality of our solutions allows one to consider many other particular cases. For example, some of the polynomial coefficients have real zeros for certain values of $n$. Taking $k^2=\tfrac14$ causes the influence of fourth-order dispersion on $\phi$ to vanish, as well as the effect of the sixth-order dispersion on the modulation frequency $B,$ and similar for eighth-order dispersion in $\phi$ when $k^2=\tfrac12$. Considering all these cases can be useful for practical application of these solutions.
\section{Conclusion}
We have presented doubly-periodic solutions of type-A and type-B for the class I infinitely extended  nonlinear Schr\"odinger equation. These solutions are expressed in terms of Jacobi elliptic functions, and have two variable periods along the two axes of the system. Being rather general, they include important cases of solutions: among them, the Akhmediev breather and the soliton solution are the limiting cases when the modulus of the elliptic functions is one or zero. As another particular case, we give a periodic solution of the mKdV equation.\\\indent
As the equation under consideration has an infinite number of free parameters, this can be useful in modelling various physical problems of nonlinear wave evolution with a large degree of flexibility in choosing the parameters. The integrability of this equation allows one to write all solutions in explicit form, adding significantly more power into the analysis.
\acknowledgments
The authors gratefully acknowledge the support of the Australian 
Research Council (Discovery Project DP150102057).
  
\end{document}